\documentclass[letter]{aa}
\usepackage{txfonts}
\usepackage{natbib}
\usepackage{graphicx}
\begin{document}
\title{XMM-Newton observations of the first unidentified TeV gamma-ray source
TeV J2032+4130\thanks{Based on observations obtained with XMM-Newton, an ESA science mission with instruments and contributions directly funded by ESA Member States and NASA.}}
\author{D. Horns\inst{1} \and   
	A.I.D. Hoffmann\inst{1} \and  
    	A. Santangelo\inst{1} \and
	F.A. Aharonian\inst{2} \and
    	G.P. Rowell\inst{3} 
   \institute{Institute for Astronomy and Astrophysics T\"ubingen (IAAT)
              Sand~1, D-72076 T\"ubingen, Germany
	     \and
          Max-Planck Institut f\"ur Kernphysik (MPIK)
          P.O. Box 10\,39\,80, D-69117 Heidelberg, Germany 
	  \and
	  School of Chemistry and Physics, University of Adelaide, Australia
	  } 
	  }
   \date{Received / Accepted }
   \abstract
   {
    The first unidentified very high energy gamma ray source (TeV J2032+4130) in the
    Cygnus region has been the subject of intensive search for a counterpart source at other
    wavelengths. In particular, observations in radio and X-rays are important to trace a population
    of non-thermal electrons.   }
   {
     A deep ($\approx 50$~ksec) exposure of TeV~J2032+4130 with \textit{XMM-Newton} has been obtained. 
     The large collection area and the field of view of the X-ray telescopes on-board of \textit{XMM-Newton}
     allow to search for faint extended X-ray emission possibly linked to TeV~J2032+4130.
   }
   {
    The contribution of point sources to the observed X-ray emission from TeV~J2032+4130 
     is subtracted from the data. 
     The point-source subtracted X-ray data are analyzed using blank sky exposures and regions 
     adjacent to the position of TeV~J2032+4130 
     in the field of view covered by the XMM-Newton telescopes to search for diffuse X-ray emission. 
   }
   {
     An extended X-ray emission region with a full width half maximum (FWHM) size of $\approx 12$~arc min
     is found. The centroid of the emission is co-located with the position of TeV~J2032+4130. The angular
     extension of the X-ray emission region is slightly smaller than the angular size of TeV~J2032+4130 
     (FWHM=$14\pm 3$~arc min).  The energy spectrum of the emission coinciding with the position and extension 
     of TeV~J2032+4130 can be modeled
     by a power-law model with a photon index $\Gamma=1.5\pm0.2_\mathrm{stat}\pm0.3_\mathrm{sys}$ and 
     an energy flux integrated between 2 and 10~keV of 
     $f_{2-10~\mathrm{keV}} \approx 7\cdot 10^{-13}$~ergs/(cm$^2$ s) 
     which is lower than the very high energy gamma-ray flux observed from TeV~J2032+4130. 
     {The energy flux detected from the extended emission region is about a factor of two 
     smaller than the summed contribution of the point sources present.}
     The energy spectrum can also be fit with a thermal emission model from an ionized plasma with a temperature
     $k_BT\approx 10$~keV.
   }
   {
     We conclude that the faint extended X-ray emission discovered in this observation is the X-ray
     counterpart of TeV~J2032+4130. {Formally, it can not be excluded that 
     the extended emission is due to an unrelated population of faint, hot ($k_BT\approx 10$~keV)  
     unresolved point-sources which by chance coincides with the position and extension of TeV~J2032+4130.} 
     We discuss our findings in the frame of both hadronic and 
     leptonic gamma-ray production scenarios. 
   }
\authorrunning{Horns et al.}
\titlerunning{XMM-Newton observation of TeV J2032+4130}
\maketitle
%
\section{Introduction}
%
Ground based air Cherenkov telescopes have detected in the last 6 years a number of unidentified
sources of very high energy (VHE, $E>100$~GeV) gamma-rays located in the Galactic plane. 
The first representative of these sources {(TeV~J2032+4130)} was discovered serendipitously with the HEGRA telescope system 
in the Cygnus region \citep{discovery,hegra1}.  

In a subsequent deep exposure for a total of  $\approx 280$~hours of observation time the source
{detection} was confirmed and 
{its properties were} studied in detail \citep{chicago,hegra2}. \\
A remarkable feature of TeV~J2032+4130 is the fact that the angular extension of the source is 
$(6.2 \pm 1.2_\mathrm{stat}\pm0.9_\mathrm{sys})$ arc~min
in radius which corresponds to a FWHM=14 arc~min. 
Gamma-ray emission from TeV~J2032+4130 has been confirmed 
independently by the Whipple collaboration \citep{whipple,konopelko}.
A gamma-ray excess from the Cygnus region albeit more extended  
was recently reported by the Milagro collaboration \citep{jordan}. 
The properties of TeV~J2032+4130 are common to most of the other unidentified VHE gamma-ray sources 
that have been discovered with the H.E.S.S. telescopes so far \citep{hess1,hess2}: 
TeV~J2032+4130 can be considered as the proto-type of the unidentified
VHE gamma-ray sources known today.  \\
%
%
So far, a number of point-like or moderately extended candidates for counterparts 
of TeV~J2032+4130  have been identified at radio \citep{paredes, butt3} and X-rays
\citep{butt1,butt2,reshmee1,reshmee2}.\\
{The origin and nature of TeV~J2032+4130 remains unclear}. 
It has been suggested that the nearby ($d=1.7$~kpc) massive stellar OB association Cyg 
OB2 (see e.g. \citet{knoedel}) 
could be an accelerator of charged cosmic rays and consequently a site of 
gamma-ray production \citep{hegra1,butt2,hegrawind}.
{This scenario has been strengthened by the recently discovered spatially extended VHE gamma-ray emission 
from the direction of the open stellar cluster Westerlund 2 \citep{reimer_hess}.}\\
Other possible explanations for the nature of TeV~J2032+4130 
have been brought forward including gamma-ray production in possible jet lobes of Cyg X-3 \citep{hegra1},
an unknown pulsar wind nebula \citep{bednarek}, or even extra-galactic source candidates as suggested {by} 
\citet{reshmee1} {and} \citet{butt3}.\\
Independent of the nature of the source, 
two different gamma-ray production mechanisms are {generally} considered. 
In the \textit{hadronic scenario}, gamma-rays are 
produced mainly via inelastic scattering of accelerated nuclei 
with the ambient medium (see e.g. \citet{hegra2,reimer,domingo}).
Alternatively, in the  \textit{leptonic} scenario gamma-rays are produced 
via inverse Compton scattering of ambient photons from a population of energetic electrons \citep{hegra2,reimer}.
Recently, {it has been suggested, that} excitation of giant dipole resonances of relativistic heavy nuclei 
in radiation dominated environments
{could be responsible for} gamma-ray production in Cyg OB2 \citep{anchordoq}.\\
While the gamma-ray observations so far have been inconclusive with respect to the 
origin of the observed signal, X-ray observations can provide additional information to identify 
the origin of the emission and to discern between the two proposed scenarios. \\
In this \textit{Letter}, we report the detection of spatially extended X-ray emission coinciding with the position
of TeV~J2032+4130.
\section{Observations and Data analyses}
%
The data were taken during two separate pointings of \textit{XMM-Newton}. 
Table~\ref{table1} summarizes the available data and configurations 
of the instruments used. 
The data were screened for soft proton flares following the method suggested in \citet{read}. 
The detectors performed almost nominally: CCD~\#6 of the MOS~1 camera had been switched off 
and CCD~\#5 of MOS2 shows an increased instrumental background below 1~keV which is two times 
larger than the background seen in the adjacent CCDs~\#4 and \#6. 
However, above 1~keV the background rate in CCD~\#5 appeared consistent with the other detectors. 
The data suffer from contamination of single scattering events from the bright X-ray source Cyg~X-3.
The contamination is most prominent in the energy band above $5$~keV. 
For this reason, the analyses {that} have been performed 
{are} constrained on the energy range from
1--5~keV where the instrumental background and stray light contamination from Cyg~X-3 are minimal. 
{A closer inspection of the contamination using ray-tracing simulations has shown that the
contribution of scattered light from Cyg~X-3 to the X-ray emission seen from TeV~J2032+4130 is negligible}.
At the same time, the vignetting of the telescopes is smallest in this energy range \citep{kirsch,pradas}.\\
Data reduction and analyses were performed using the Standard XMM-Newton Science Analysis Software \textit{SAS} v7.0.
\begin{figure*}[t]
\mbox{
 \includegraphics[height=6.5cm]{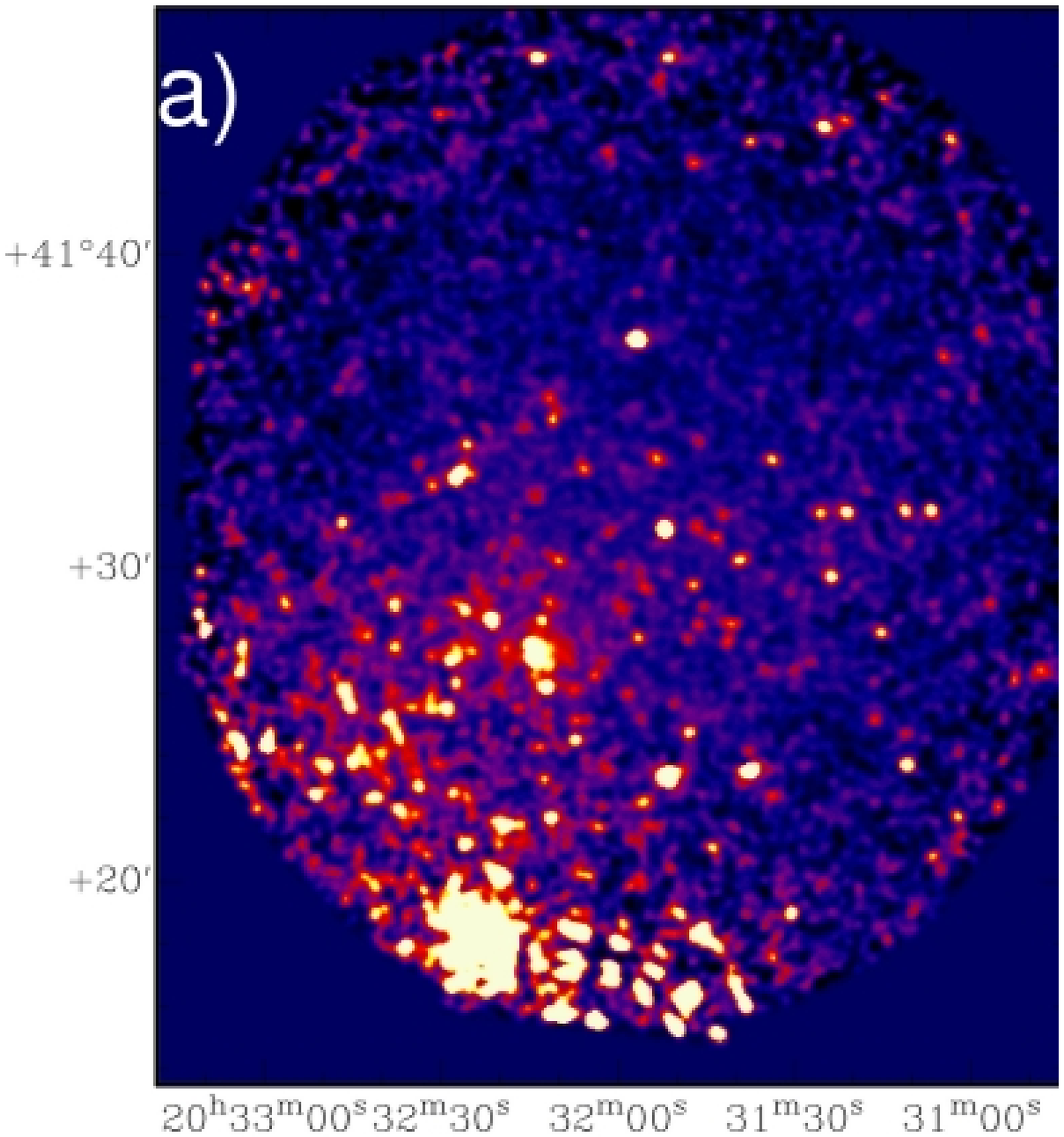}
 \includegraphics[height=6.5cm]{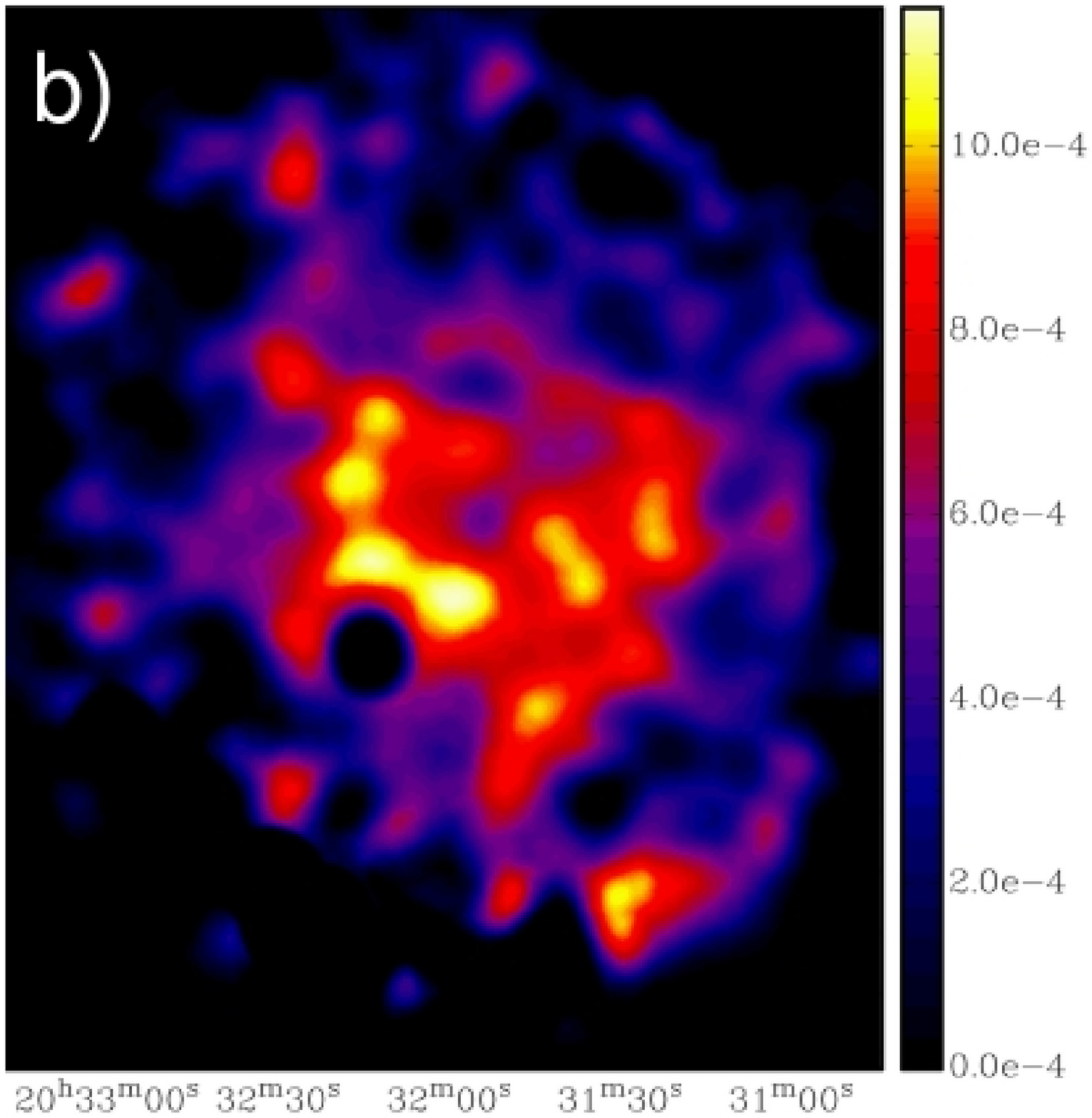}
 \includegraphics[height=6.5cm]{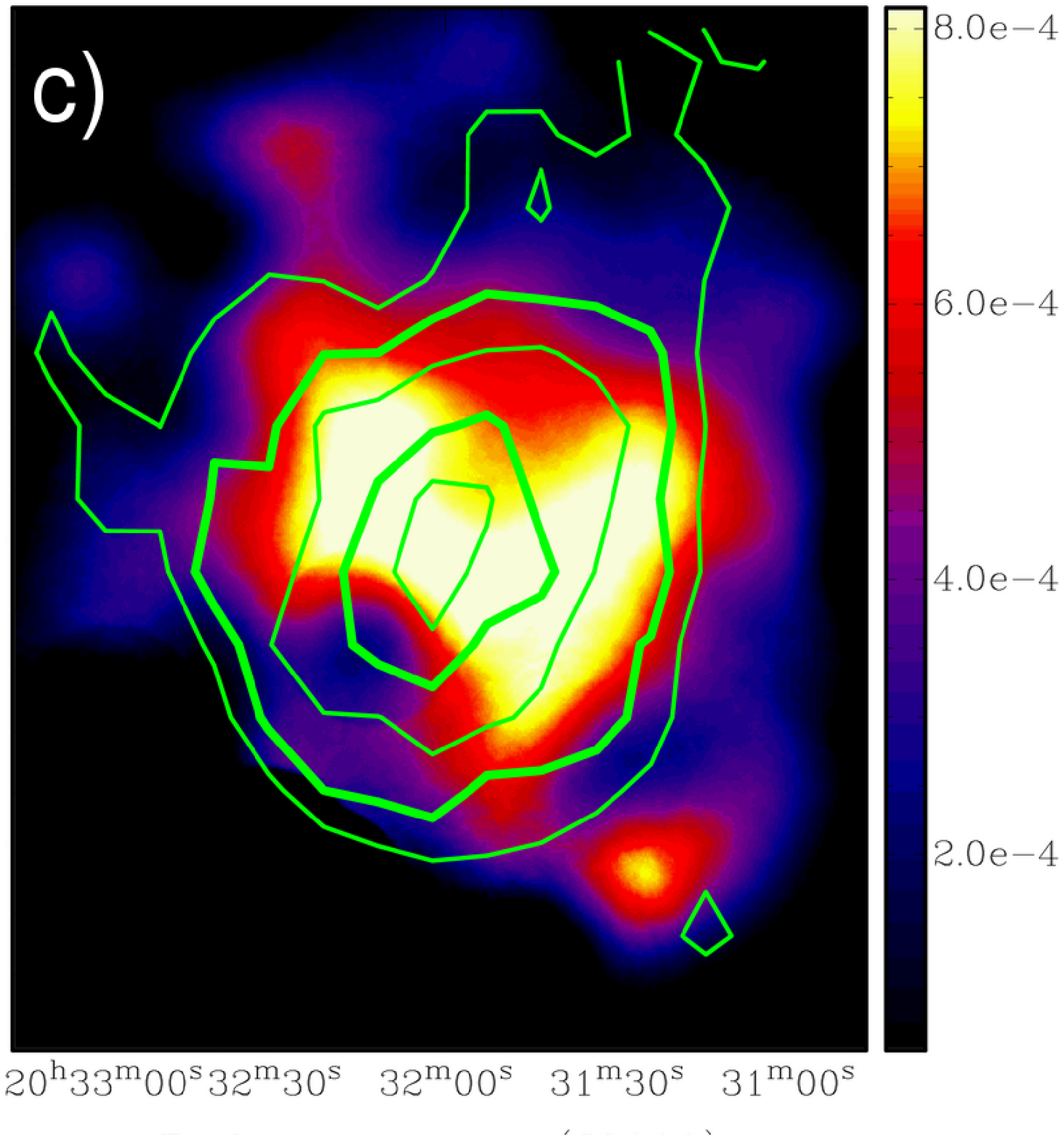}
 }
  \caption{\label{mosaic} 
  {Background subtracted and exposure corrected images obtained from the combined
  exposures of the MOS1 and MOS2 cameras (in units of counts/(sec arcmin$^2$). While the left image shows the 
  emission including all point sources, the middle image  shows the
  image after subtracting off all point sources and smoothed with 
  a Gaussion of 45 arc~sec width}. Finally, the right image is obtained by
  smoothing the source subtracted image with a Gaussian of 1.5 arc min width and changing the scaling to highlight the extended
  X-ray emission present.
  The green contours indicate (in linear spacing) the
  significance contour of the HEGRA observations (starting at 3~$\sigma$). 
  }
\end{figure*}
%
\subsection{Search for diffuse X-ray emission}
In an initial step of the data analyses, a catalogue of sources detected in the 1-5~keV energy range 
is assembled using the \texttt{edetect\_chain} task simultaneously for the MOS1, MOS2, and EPIC pn  
cameras for each of the two exposures. {The minimum detection likelihood is set to a low value of 2. 
This way, 13.5~\% of the detected sources are due to statistical fluctuations while at the same time, 
more and fainter sources are detected than with the default value of 10.}
A detailed study of the point sources detected in this XMM-Newton observation will be presented elsewhere. 
Here, the source catalogue has been used to {generate model images of the point sources separately 
for each camera and pointing taking into account the position dependent point spread function of 
the XMM-Newton mirrors. 
These model images are then subtracted off
the individual frames of the three cameras and for the two pointings.
Within the 6.2$'$ region covered by the extension of the
TeV source, a total energy flux of 
$f_{1-5~\mathrm{keV}}\approx 10^{-12}$~ergs/(cm$^2$~s) associated with point sources 
is observed.
}
In order to 
{subtract the particle and extragalactic photon background
blank sky event files for the corresponding observation modes and filter settings have been obtained from 
the XMM guest observers' facility\footnote{\texttt{ftp://xmm.esac.esa.int}}. The blank field data have been 
processed to match the observational data and the resulting images are subtracted off the pointings towards TeV~J2032+4130.}  %
The images for the MOS1 and MOS2 cameras obtained from the two exposures are combined to form a background subtracted mosaic. 
{The images of the EPIC pn camera have been omitted because of artefacts resulting from over-subtracting
the contribution of point sources. For each of the four remaining images, exposure maps have been generated
using the SAS task {eexpmap}.
The exposure maps are combined in a mosaic which is then trimmed
to a minimum exposure of 5~ksec corresponding to roughly 5~\% of the peak value. The background
subtracted mosaic is then divided by the exposure map to obtain an exposure corrected, background subtracted,
 and source free image.}
{Fig.~\ref{mosaic} shows three images of the combined exposures of the MOS1 and MOS2 cameras before
the subtraction of the model images of the point sources (a), after the subtraction
{and smoothing with Gaussian of fixed width of 45 arc~sec} (b), and finally,
the image after convolving it with a Gaussian of fixed width of 1.5 arc~min (c). Note the presence of
scattered light along ring segments in the southern part of Fig.~1a.} 
{While Fig.~1b shows indications for the presence of a rim-like feature resembling
the morphology observed in radio by \citet{paredes}, it should be noted
that the subtraction of point-sources has an impact on the observed morphology on angular
scales small compared to the extension of the detected extended emission.}
For a comparison with the extension and morphology of the gamma-ray
source TeV~J2032+4130, contours are overlaid to {Fig.~\ref{mosaic}c}. 
The X-ray image shows an extended emission region which is co-located with 
TeV~J2032+4130 and of similar extension. 
{The exposure corrected image of the diffuse emission is fit by a Gaussian 
with a width of $(5.1\pm0.3)$~arc~min which corresponds to 
a FWHM$=(11.7\pm 0.6)$~arc~min\footnote{all errors given throughout the paper are to the 
confidence of one standard deviation}.} 
\subsection{Energy spectrum of the diffuse X-ray emission} 
{In order to determine the energy spectrum of the diffuse X-ray emission, the contribution of
the point sources has to be removed by excising regions around the sources. We use an excising region adapted
to the point spread function and brightness of the source. The size of the region is chosen such that the
relative contribution of the source to the local background is less than 20~\% at the boundary of the
excised region. The wings of the point spread function lead to a contamination of the detected diffuse emission
which amounts to 27~\% of the excess signal seen. This effect of contamination would be however compensated by
the opposite effect caused by the diffuse emission present in the excised region ($\approx 25$~\% )
and therefore excluded from the reconstructed energy spectrum. We estimate the systematic uncertainty 
of this contamination to be 10~\% of the total flux.}
 The energy spectra of a source region with a radius of $6.2'$ centered on the position of TeV~J2032+4130  
 from the MOS2 and EPIC~pn cameras  of the second pointing (ObsID 0305560201) are extracted. 
 The second pointing is centered 5$'$ north
 of the centroid of TeV~J2032+4130 and therefore, we can choose a background region which is 
 mirrored through the center of the field of view and 
 has a similar acceptance as the source extraction region. 
 Using the background region in the same field of view, the
 background estimate includes all relevant background components including X-ray emission from the Galactic ridge and
 the particle and instrumental background specific to this observation. 
 The small azimuthal modulation of the acceptance of the MOS cameras can be neglected 
 along this direction. 
 The MOS1 camera is not used because the disabled CCD\#6 unfortunately  does not allow to cover the entire source.\\
 The response files are generated for these particular observations using 
 the standard tools \texttt{arfgen} and \texttt{rmfgen}. Finally, the 
 geometrical areas of the source and background regions are calculated  taking gaps between the CCDs, bad pixels, 
 and point source exclusion regions into account. 
 The resulting background subtracted X-ray spectrum is fit with a power-law as well as 
 an optically thin hot plasma model (\texttt{apec}) including photoelectric absorption (\texttt{phabs}). 
 The spectral fitting of the two spectra was done using the
\texttt{xspec} v11.3.2p spectral fitting package \citep{arnaud}.  
 The fit describes the data well (see Table~\ref{table2} for
 a summary of the fit parameters; only statistical $1~\sigma$ errors are quoted).\\
{In order  to estimate the influence of faint, 
unresolved point sources present in the stellar cluster, 
we have reduced X-ray imaging data taken with the Chandra satellite (Obs.\# 4501 for 48.6 ksec
on TeV J2032+4130 and Obs.\# 4511  for 98.7 ksec in the south-west of TeV J2032+4130, centered on the 
core of Cyg OB2). Based upon a comparison of the Chandra and XMM-Newton source catalogues 
and the study of the X-ray source population present in the 
Cyg OB2 cluster \citep{colombo}, we estimate  that  point sources below the XMM-Newton
detection limit can contribute up to $\approx 30$~\% of the observed diffuse emission. It is however not
straight-forward to correct the observed excess for the unresolved point sources detected with Chandra as 
 the flux of the sources is generally found to vary with time.
We estimate the systematic uncertainty on the flux conservatively 
to be 50~\% combining the influence of the unresolved
point sources ($\approx 30$~\%), uncertainties on the background subtraction ($\approx 10$~\%), as well as 
the tails of the point spread functions of excised point sources ($\approx 10$~\%).
The total, unabsorbed
 energy flux derived from the model fit  (see Table~\ref{table2})
 is found to be $f_{2-10}=(7.3\pm3.5_\mathrm{sys}\pm1.1_\mathrm{stat})\times 10^{-13}$~ergs/(cm$^2$~s). 
}
\begin{table}
 \caption{\label{table1} Summary of observation times and configurations on the
 target ``TeV~J2032+4130''. All instruments were
 operated in full frame mode with a medium filter. The exposure quoted in parenthesis
 is for the EPIC pn camera.}
 \begin{tabular}{lcccc}
 \hline 
  ObsID  & Date   & R.A.    & Dec.                 & Exposure  \\
         &        & J2000   & J2000  & [ksec]   \\
  \hline 
  \hline
0305560101  & 2005-10-21 & 20:31:57  & 41:29:58 & 27.3(23.6)  \\
0305560201  & 2005-10-25 & 20:31:57  & 41:34:55 & 25.8(20.5)  \\
\hline
 \end{tabular}
\end{table}
\begin{table}
 \caption{\label{table2} Summary of the fit results of the X-ray energy spectrum
 with a power-law model (\texttt{powerlaw})  or thermal emission (\texttt{apec} 
 with fixed solar abundances of the plasma) 
 including photo-electric absorption (\texttt{phabs}). }
 \begin{center}
 \begin{tabular}{lrr}
 \hline
 Parameter                        & Value (powerlaw)      & Value  (apec)\\
 \hline \hline
  $N_H$  $[10^{21}$~cm$^{-2} ]$   & $3.5 \pm 1.6$         & $3.2\pm 1.1$ \\
  $\Gamma$, $k_B T$ $[$keV$]$     & $1.5 \pm 0.2$         & $10.5\pm 3.2$\\
  $f_{2-10~\mathrm{keV}}$ $[10^{-13}$~ergs/(cm$^2$ s)$]$  & $7.3\pm 1.1$                   & $7.4\pm1.1$    \\
  $\chi_\mathrm{red}^2$ (d.o.f.) & 1.02 (862)         & 1.01 (861)\\
\hline
 \end{tabular}
 \end{center}
\end{table}
\begin{figure}[t!]
\includegraphics[width=0.98\linewidth]{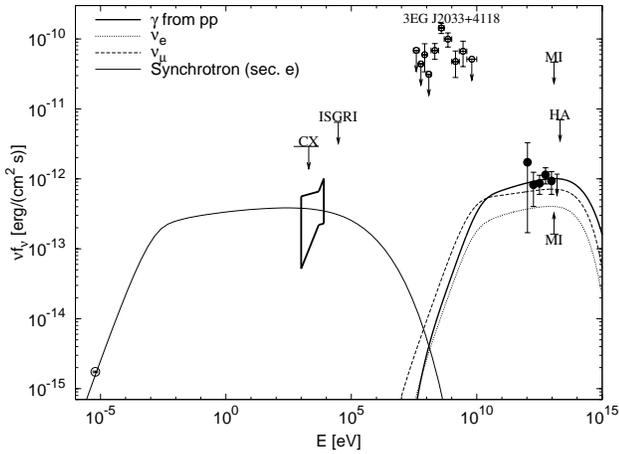}
\caption{The spectral energy distribution of TeV~J2032+4130: 
the bow-tie indicates the energy spectrum as measured with 
XMM-Newton including systematic and statistical uncertainties. The lines indicate the result of a model calculation 
for a hadronic gamma-ray and neutrino production scenario for a {continuously active accelerator of protons 
up to a maximum energy of 5~PeV at an} age {of} $t_\mathrm{age}=2\,500$~yrs and magnetic field
$B=1~$mG.  
The radio flux point of a possible non-thermal
extended radio-source at $\lambda=20$~cm \citep{paredes} 
is assumed to be an upper limit to the actual radio emission associated with
TeV~J2032+4130.
The upper limits between 0.5--5~keV (CX, Chandra) and 20--40~keV (ISGRI, INTEGRAL) are taken from \citet{butt2}. The energy spectrum of the 
EGRET source 3EG~J2030+4118 \citep{hartman} 
is considered to be an upper limit. 
The Milagro (MI) upper limit
is the integrated emission in a 3x3 $(^\circ)^2$ region \citep{jordan}
while the lower limit (MI) is scaled to the solid angle
covered by TeV~J2032+4130 assuming a uniform surface brightness.
The upper limit from the HEGRA-AIROBICC (HA) wide angle Cherenkov
detector is taken from \citet{hegra1}. \label{sed1}}
\end{figure}
\begin{figure}[t!]
\includegraphics[width=0.98\linewidth]{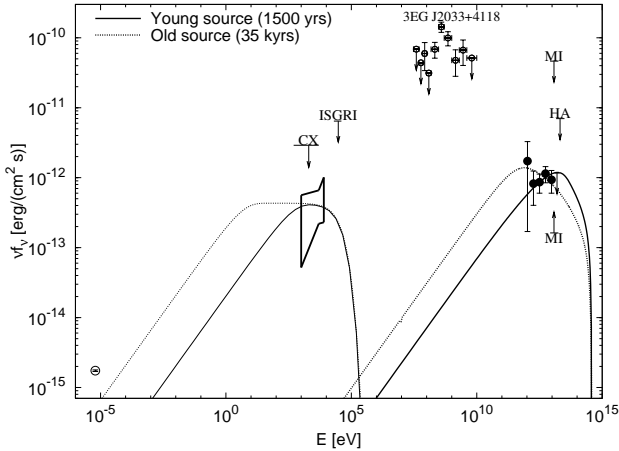}
\caption{In the leptonic scenario, the data are described well by a 
young source of an age of 1500~yrs, 
a  magnetic field of $B=3~\mu$G and an energy density of the seed photon field of $w_\mathrm{IR}=3$~eV/cm$^3$ 
with a grey body temperature of  $T=10~$K. As
an alternative, an older source at an age of 35~kyrs is shown ($B=3~\mu$G, $w_\mathrm{IR}=1$~eV/cm$^3$). For a description of the 
multi-wavelength data see the caption of Fig.~\ref{sed1}. \label{sed2}}
\end{figure}
\section{Conclusions}
The XMM-Newton observations presented here indicate the presence of an extended (FWHM=11.7 arc~min) X-ray source co-located
with the first unidentified VHE gamma-ray source TeV~J2032+4130 discovered with the HEGRA air Cherenkov telescope system. 
The size of the X-ray source is similar to the one of TeV~J2032+4130. The energy spectrum can be fit by a power-law model or
by a thermal emission model with a plasma temperature of $k_BT\approx 10$~keV. 
 The unabsorbed energy flux of the X-ray source in the energy range from 2--10 ~keV is a factor of 2--3 smaller  
 than the one observed from TeV~J2032+4130 at energies from 1--10~TeV. \\
 We note  that the observed extended X-ray emission could in principle be a so far unknown population of faint X-ray sources
that are by chance distributed at an angular size which is similar to the one of TeV~J2032+4130. The energy spectrum of these
sources would have to be different from the bulk of the stellar X-ray sources detected from Cyg OB2 \citep{colombo} 
which on average show a thermal spectrum with $k_BT\approx 1\ldots 3$~keV. 
Taking these considerations into account, we conclude that the observed extended
X-ray emission is the X-ray counterpart of TeV~J2032+4130. \\
 Initial modeling of the X-ray and gamma-ray energy spectra {using a hadronic gamma-ray production scenario} indicates that
 the observed spectra can be naturally explained by a young (a few kyrs) 
 ``Pevatron'' accelerator. {The observation of synchrotron X-ray emission 
 up to $\approx 5$~keV constrains 
 the product of the square of maximum energy of accelerated protons 
 ($E_\mathrm{max}^2$) and magnetic field ($B$) to 
 exceed $E_\mathrm{max}^2\cdot B\ga 5~\mathrm{PeV}^2~\mathrm{mG}$}.
 In this picture, the
 gamma-ray energy spectra would continue without cut-off well beyond 10~TeV. The hard X-ray emission from a
 similar source size region as the gamma-ray emission is a natural prediction within this 
 model where the X-rays are produced by synchrotron emission of secondary electrons. Notably, in the frame of this model, 
 the extended gamma-ray emission found 
 by the Milagro collaboration would be explained by gamma-rays produced by accelerated particles of energies $\ga 100$~TeV which have already escaped
 the accelerator.\\
A leptonic scenario provides a valid explanation of the observations as well (see Fig.~\ref{sed2}). 
In this scenario, a young accelerator with a low magnetic field of a few $\mu$G accelerating electrons {following
a power-law distribution with $dN/dE \propto E^{-2}$} reaching up to 
energies of a few 100 TeV can provide a good fit to the X-ray and gamma-ray data. In contrary to the hadronic scenario, 
 the gamma-ray energy spectrum above 10~TeV is expected to be rather soft due to un-avoidable Klein-Nishina suppression 
 of inverse Compton scattering. Furthermore, the seed photon density has to be higher 
 than the average value in the interstellar medium (which is not unlikely given the possible proximity to 
 the Cyg OB2 stellar association and the large stellar extinction towards that region indicating high density of dust). 
 In the leptonic scenario, the X-ray spectrum can be expected to vary at different
 parts of the source due to cooling effects. More X-ray observations will be required to detect spectral variability.\\
 Finally, neutrino observations will prove decisive to discern {between} the two emission scenarios
 {(since neutrinos are only expected in the hadronic scenario)}. Even though the neutrino flux calculated here 
 is very likely not detectable with the coming generation of neutrino telescopes like IceCube \citep{icecube}, it may be that
 the gamma-ray flux detected with Milagro from the same region is tracing the high energy particles which have already left the accelerator
 and fill a much larger volume. In this case, the region \textit{around} TeV~J2032+4130 would be a powerful high energy neutrino source and
 would be detectable with future neutrino telescopes \citep{beacom}.\\
 It should be noted that the observations of TeV~J2032+4130 carried out with XMM-Newton are so far the deepest observations
 available for any of the unidentified {VHE} gamma-ray sources. It will be interesting to 
 perform similar observations to search for faint, extended X-ray counterparts of other unidentified VHE gamma-ray sources. 
\begin{acknowledgements}
We acknowledge  the support of the Deutsches Zentrum f\"ur Luft- und Raumfahrt under grant number 50OR0302 and the support of the
Eberhard Karls Universit\"at T\"ubingen. We wish to thank Olaf Reimer and Facundo Albacete for providing the source lists found in the Chandra 
observations and  Josep Paredes for making their results available prior to publication. 
This research has made use of NASA's Astrophysics Data System. We thank the anonymous referee for
valuable comments.
\end{acknowledgements}

\end{document}